# Optical centers in cubic boron nitride and diamond: remarkable similarities

Konstantin Iakoubovskii[a,*], Andrey Katrusha[b], Weihua Peng[b], Jianguo Peng[b]

[a] *Central European Institute of Technology, Brno University of Technology, Purkyňova 123, 61200 Brno, Czech Republic*
[b] *Jilin Diamond Technology Research & Development, Jianan Road 177, Luyuan District, Changchun, China*

**Abstract**
We present a comparative study of optical absorption and luminescence from cubic boron nitride (cBN) and diamond grown by the high-pressure high-temperature technique in the same cubic press. We note remarkable similarities in spectral and spatial dependences for these two materials. Using the previous identification of defects in diamond, we tentatively assign the optical center responsible for yellow color in some cBN crystals to substitutional oxygen at the nitrogen site ($O_N$), the RC1 and RC3 centers to a defect comprising $O_N$ and a boron vacancy ($V_B$) in the neutral and negative charge states, respectively, the GC1 center to a nickel-related defect, the 1.816 eV (683 nm) luminescence peak to a $Si_N$-$V_B$ complex, and the BN1 center to an interstitial-related defect.

*Corresponding author: iakoubovskii@vut.cz

## 1. Introduction

Cubic boron nitride (cBN) and diamond share many exceptional physical properties—wide band gaps, high thermal conductivity, and the zinc-blende lattice—yet their defect identification and control are at very different stages of maturity. While diamond hosts a large catalog of well-identified optical centers, the microscopic origin of luminescence centers in cBN remains unresolved. Even the bandgap value $E_g$ of cBN is still debated: optical measurements suggest $E_g \sim 6.4$ eV while electron energy loss spectroscopy yields $E_g \sim 10.1$ eV [1], and this disparity could be attributed either to the residual defect-related levels inside the gap [1] or simply to different gaps being studied – direct or indirect [2]. For comparison, the values of indirect and direct gaps in diamond are 5.5 and 7.2 eV, respectively [3].

Nevertheless, the similarity in properties of cBN and diamond suggests an opportunity: the extensive knowledge on defect assignment in diamond may provide a roadmap for interpreting cBN optical spectra.

Table 1. Comparison of basic properties of c-BN and diamond [1,4,5]: average atomic number and mass *A* and *Z*, lattice constant *a*, refractive index *n*, static dielectric constant ε, maximum phonon energy $E_{Ph}$, band gap $E_g$, bulk modulus *K*.

| Material | Z | A | a (nm) | n | ε | $E_g$ (eV) | $E_{Ph}$ (meV) | K (GPa) |
|---|---|---|---|---|---|---|---|---|
| cBN | 6 | 12.0 | 0.3617 | 2.1 | 7.1 | 6-10 | 162 | 400 |
| Diamond | 6 | 12.4 | 0.3567 | 2.4 | 5.7 | 5.5 | 165 | 440 |

While the defect physics of diamond is now well developed, with many optical centers unambiguously identified [6], the situation in cBN remains far less mature. More than a dozen luminescence centers have been reported in cBN, often labeled according to their production history (GC, PC, RC), but their microscopic origin is still debated. Previous studies have linked these centers to irradiation damage, impurity incorporation [1], or structural defects, yet definitive assignments remain scarce. Even the fundamental optical gap of cBN and the nature of its dominant intragap states are still under active discussion [1].

Given the close structural and electronic parallels between cBN and diamond, it is natural to ask whether defects that are well understood in diamond may have direct analogues in cBN, shifted in energy due to the larger band gap in cBN and modified by polarity. This comparative approach has not been systematically explored, despite its potential to resolve long-standing ambiguities in cBN defect identification.

In this work, we perform a side-by-side study of optical absorption and luminescence in cBN and diamond crystals grown under very similar high-pressure high-temperature (HPHT) conditions in the same cubic press. This experimental configuration minimizes extrinsic variability and allows direct comparison of intrinsic defect signatures. We observe that many optical features in cBN have clear counterparts in diamond, including broad absorption thresholds, dislocation-related blue luminescence, vacancy-related centers, and impurity-related emission lines.

By leveraging the well-established defect assignments in diamond, we propose tentative microscopic models for several prominent cBN centers. In particular, we attribute the yellow coloration of our cBN crystals to substitutional oxygen at the nitrogen site ($O_N$), supported by elemental mapping showing oxygen enrichment in yellow regions. We further suggest that the RC1 and RC3 centers correspond to $O_N$–$V_B$ complexes in different charge states, analogous to the $NV^0$ and $NV^-$ centers in diamond. The 1.816 eV luminescence line is reassigned to a $Si_N$–$V_B$ split-vacancy defect, paralleling the $SiV^-$ center in diamond, while the BN1 center is linked to an interstitial-related complex based on its characteristic phonon coupling.

Overall, our results reveal a striking one-to-one correspondence between many optical centers in cBN and diamond, providing a coherent framework for interpreting cBN luminescence and highlighting the importance of controlled impurity incorporation in future cBN synthesis.

## 2. Experimental details

Diamond and cBN crystals were synthesized at 6.5–7.0 GPa and 1600–1800 ºC using a cubic press and the temperature gradient method. We used a Ni-based or Fe-based solvent catalyst and graphite or hexagonal boron nitride (hBN) as the source material. Silicon doping was achieved by introducing 1wt% of silicon powder to the growth medium, while radiation defects were generated by 2 MeV electrons at a dose of $10^{18}$ e/cm$^2$. The as-grown crystals, with sizes as large as 8.5 mm, were thoroughly cleaned in mixtures of sulfuric, nitric and hydrochloric acids. Six samples from each growth run were selected for characterization, and the presented results were reproduced on at least five specimens for each batch.

Luminescence and Raman spectra were measured in backscatter geometry with a WITec alpha 300R confocal microscopy setup equipped with 633 nm (HeNe), 532 nm and 355 nm (Nd:YAG) linearly polarized lasers. Chemical composition was assessed by energy-dispersive X-ray spectroscopy (EDS, 30 kV) inside an Amber X2 scanning electron microscope (Tescan) and by X-ray fluorescence (XRF, Bruker Tornado M4). Optical absorption was measured with a Jasco V770 spectrometer, with the required region of the sample selected with a metal mask.

## 3. Results and discussion

3.1. Yellow color

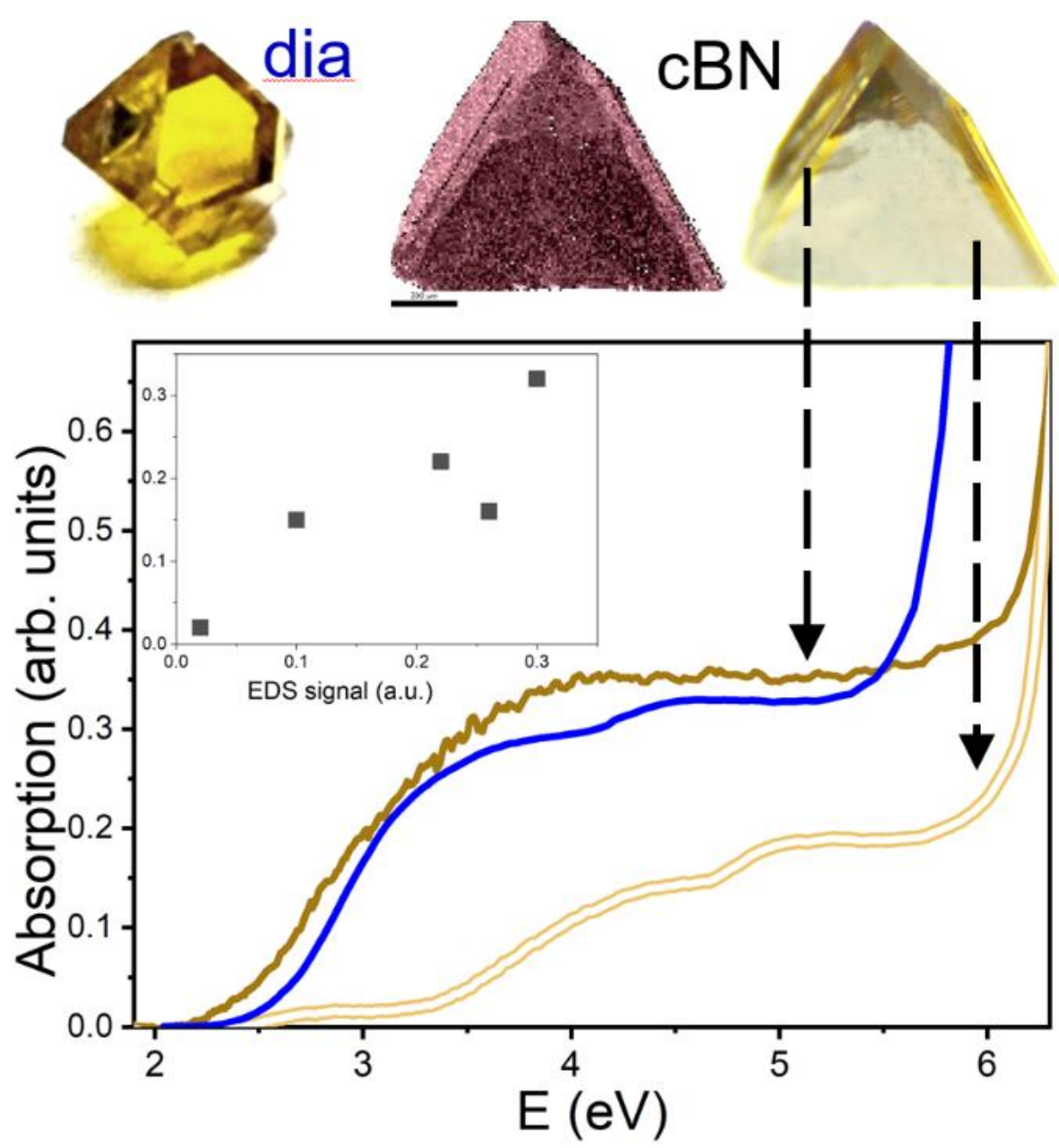

*Figure 1. Room-temperature optical absorption spectra of HPHT cBN and diamond crystals. Black-white and dark yellow curves correspond to white and yellow regions in the cBN crystal shown in the top-right photograph. The diamond spectrum is shown in blue. The top-center figure is the EDS map of oxygen in the same sample (scale bar 0.2 mm). The inset correlates the optical absorption at 3 eV with EDS signal.*

Most of our cBN and diamond crystals were yellow in color with accidental white zones that were assigned to the (100) growth sectors. Optical absorption measurements on the yellow regions revealed a featureless and broad absorption threshold that starts at ca. 2 eV and flattens only at ca. 3.5 eV (see Figure 1). This threshold is well known in diamond and is assigned to electron transitions from the substitutional nitrogen ($N_S$) donor to the conduction band [7]. This assignment essentially hinges on the identification of isolated nitrogen as the major impurity in HPHT diamond. Our XRF and EDS measurements reveal that such major impurity in our cBN samples is oxygen, which can reach concentrations of ca. 0.1at%. Mapping of the oxygen distribution revealed its elevated presence in the yellow zones of cBN (see EDS map in the center top of Figure 1).

Indeed, theoretical modeling suggests that oxygen, as an electronegative atom, should substitute for nitrogen in the cBN lattice (hence its $O_N$ labeling) and form a deep donor with an activation energy of ~1.8 eV [8,9,10], which is isoelectronic to $N_S$ in diamond. Meanwhile, experiment reveals n-type conductivity in nominally undoped HPHT cBN crystals. Oxygen was the major impurity in those crystals, with concentrations ranging from tens to hundreds parts per million (ppm) depending on color [11,12]. Note that the theoretical prediction of $O_N$ as the most stable oxygen form also agrees with the reported X-ray absorption near edge structure spectra [13].

Hence, we tentatively attribute the yellow color and the corresponding broad threshold in absorption spectra of cBN to the individual oxygen atom that substitutes for nitrogen.

3.2. The blue luminescence band

Most diamonds exhibit broad luminescence bands centered at 2.8 eV (see black spectrum in Fig. 2c). Based on cathodoluminescence imaging, it was assigned to an intrinsic structural defect located at dislocations [14-17]. A very similar band appears in cBN, as shown by the literature survey [11,18-21] and the red spectrum in Fig. 2c. Our spatial mapping results summarized in Fig. 2 reveal that this band is related to structural, dislocation-like defects.

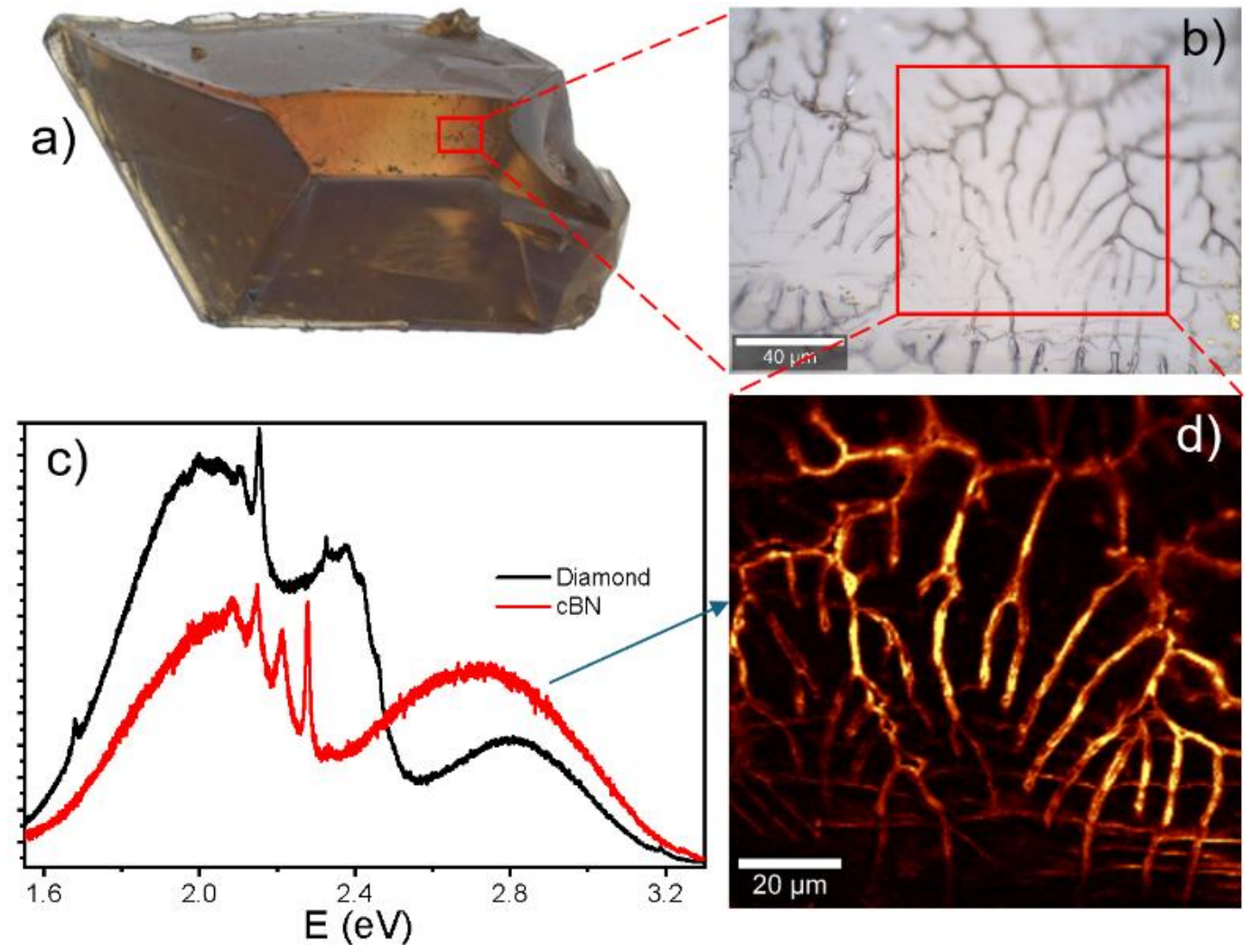

*Figure 2. (a,b) optical image of dislocation-like defects in a 2-mm-wide cBN crystal. c) comparative luminescence spectra (355 nm excitation) from HPHT diamond and cBN, showing the broad blue band centered at ca. 2.8 eV. d) map of 2.8 eV emission in the region marked by the red rectangle in panel b).*

3.3. Vacancies and RC centers

Irradiation of cBN with high-energy particles produces RC1, RC2 and RC3 luminescence centers (where RC stands for radiation cubic) with zero-phonon lines (ZPLs) at ca. 2.27, 2.15 and 1.99 eV, respectively

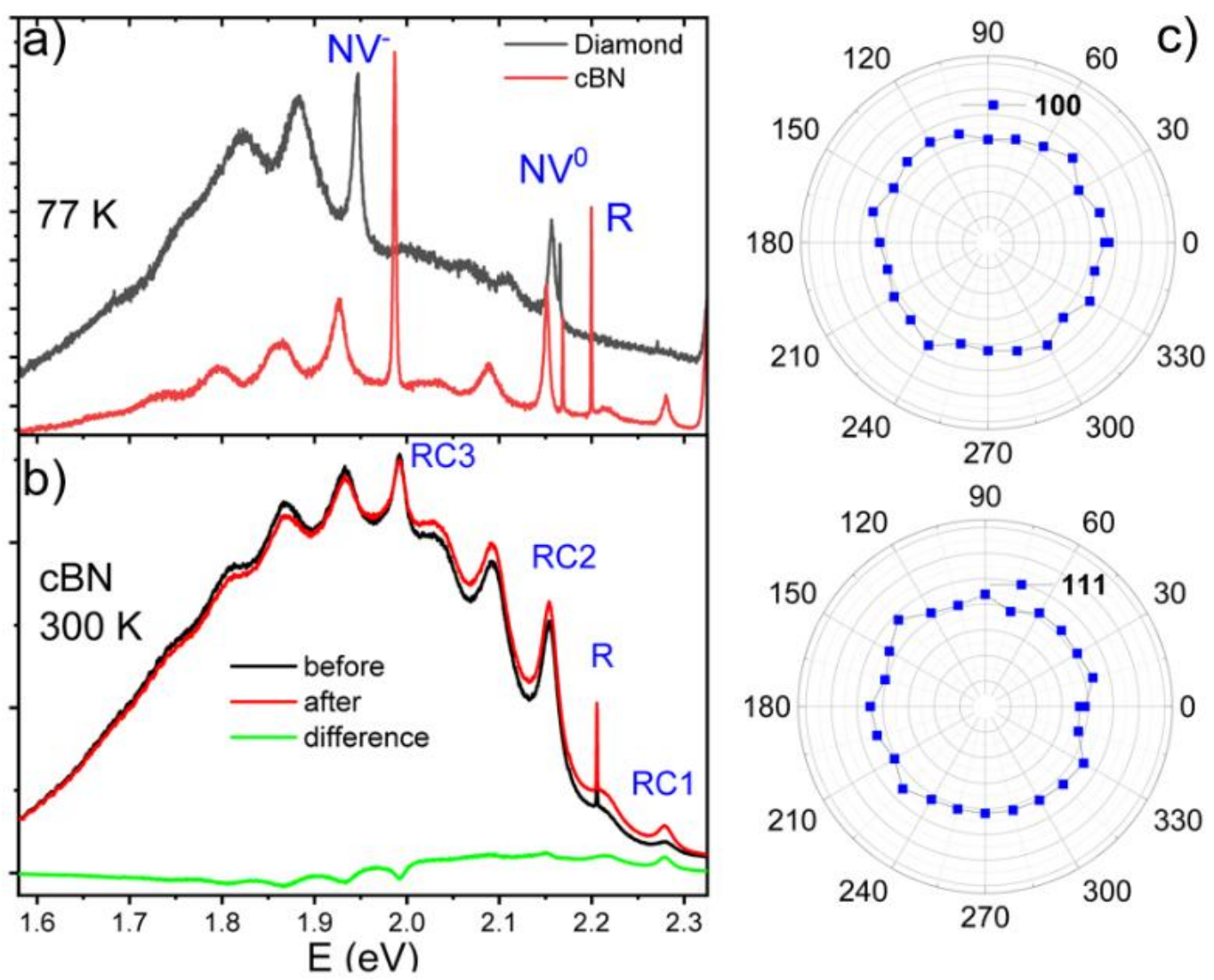

*Figure 3. a) Luminescence spectra measured from HPHT diamond and cBN at 77 K under 532 nm excitation. b) Room-temperature luminescence spectra from cBN under 532 nm excitation, before and after additional 440 nm illumination, and the difference spectrum revealing that the 440 nm irradiation quenches RC3 and enhances RC1 emission. R marks Raman peaks. c) Polarization dependences for the RC2 (same for RC1 and RC3) center for analyzer rotation in (100) and (111) plane.*

[1]. They disappear after annealing at 700-800 °C and hence were associated with vacancies [22]. These centers can also be created by high-energy laser pulses, presumedly via photoacoustic effect and concomitant plastic deformations [23].

Figure 3a compares low-temperature spectra of the RC1, RC2 and RC3 spectra in cBN with those of the neutral and negative nitrogen-vacancy centers in HPHT diamond ($NV^0$ and $NV^-$). A remarkable similarity is seen in the spectral shape, with all these centers showing a moderately strong coupling to 64-meV phonons. The ZPL positions are also similar between $NV^0$ and RC1 (2.156 and 2.281 eV) and $NV^-$ and RC3 (1.945 and 1.986 eV). Illumination of cBN with 440 nm light partly quenches the RC3 centers and enhances RC1 emission (see Fig. 3b), suggesting that they are different charge states of the same defect. The same photochromism was previously observed for $NV^0$ and $NV^-$ [24]. Furthermore, polarization dependences of RC1-3 signals (see Fig. 3c) are consistent with the trigonal symmetry. With this background, we tentatively assign the RC1 and RC3 centers to $O_NV^0$ and $O_NV^-$ defects in cBN, which are isoelectronic to the diamond's $NV^0$ and $NV^-$. These defects should be thermally stable, which explains their strong intensity in our HPHT crystals grown at temperatures as high as 1800 ºC for up to 96 hours. Subsequent annealing at 1000 ºC for 1 hour did not affect the RC1-3 intensities, confirming that these defects are stable, and suggesting that their disappearance after annealing at 800 °C in electron-irradiated cBN may be related to recombination with migrating interstitials rather than intrinsic instability of the $O_SV$ defects in cBN.

Our identification of $O_SV$ as a deep center in cBN with a ZPL at ca. 2 eV agrees with multiple theoretical studies, which state that $O_SV$ in cBN is a stable deep center with similar properties to the NV defect in diamond. Theory predicts a ZPL of 1.6 eV for $O_SV$, ~0.4-0.5 eV lower than our measurement, but it also admits underestimating the cBN bandgap by approximately the same amount [25,26].

### 3.4. GC1 and Ni-related luminescence

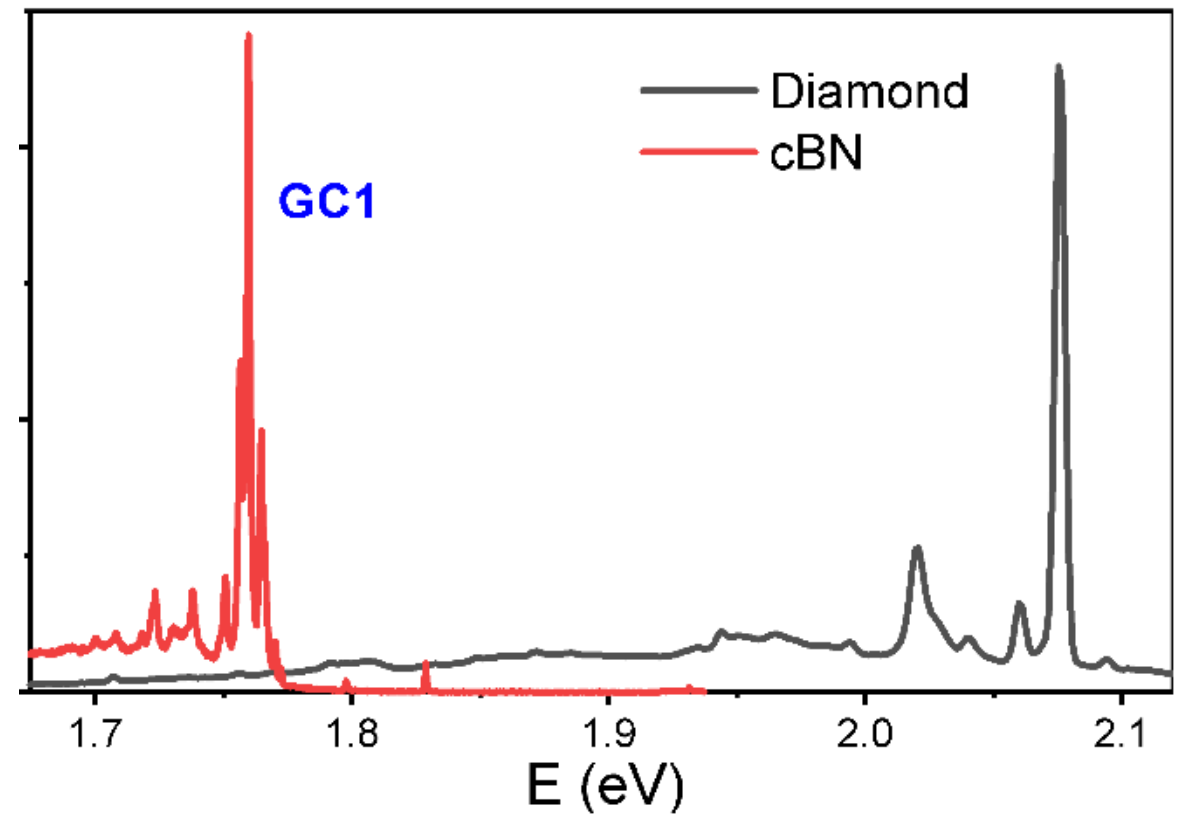

*Figure 4. Luminescence spectra (532 nm, 77 K) from HPHT cBN and diamond crystals grown with a Ni-based solvent-catalyst.*

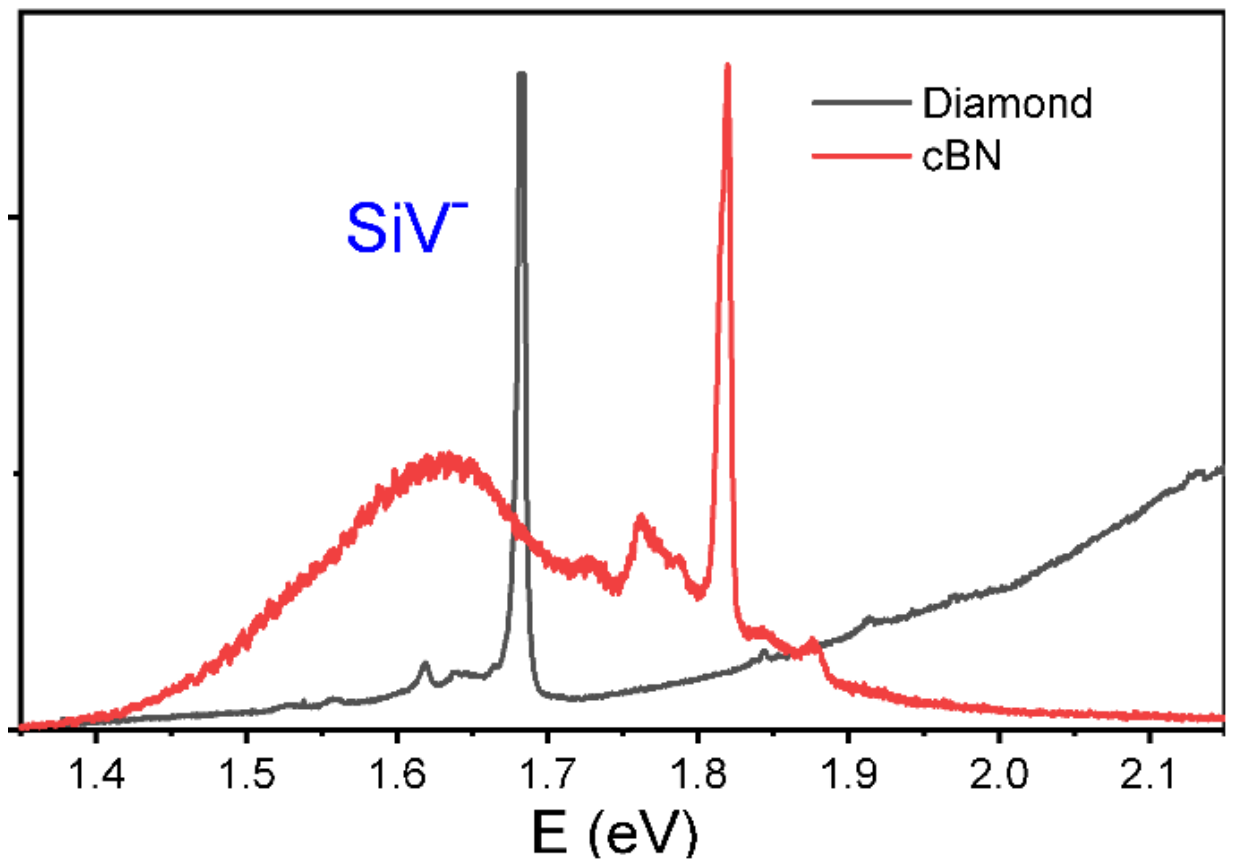

*Figure 5. Si-related luminescence peaks (532 nm, 77 K) in as-grown Si-doped HPHT diamond and cBN. The broad 1.63 eV band in cBN is unrelated to Si.*

Many as-grown cBN crystals exhibit a luminescence center with the ZPL at 1.76 eV, which was labeled as GC1 (general cubic 1) and assigned to nitrogen vacancies [1]. A luminescence spectrum of GC1 is shown by the red curve in Fig. 4. Thanks to the high quality of our HPHT crystals (Raman linewidth <2 $cm^{-1}$), we were above to resolve a rich fine structure in the GC1, which was not seen in the RC1-3 centers in the same sample. Such structure is rather common for nickel-related luminescence in diamond, where it originates from the Ni-related spin-orbit splitting [27]. A representative example is shown by the black curve in Fig. 4. Given that both cBN and diamond crystals used for Figure 4 were grown with a Ni-based catalyst, we assign the RC1 center to a Ni-related defect.

### 3.5. Si-related luminescence

When doped with Si, cBN exhibits an additional luminescence line at 1.82 eV, which was tentatively assigned to a Si atom substituting for nitrogen ($Si_N$) [28]. In this work, we have reproduced these results and produced Si-related luminescence in both cBN and diamond as shown in Fig. 5. Similar to the case of RC1-3 and NV centers, a remarkable similarity both in spectral shape and ZPL positions is seen for the 1.82 eV center in cBN and the previously identified $SiV^-$ emission in diamond [29]. Hence, we tentatively re-assign the 1.82 eV center in cBN from $Si_N$ to $Si_NV_B$ split-vacancy defect where the relatively large Si atom is shifted from the lattice site toward the vacancy thereby reducing the local strain [29].

This re-assignment agrees with experimental fact that we do not observe Si-related emission in millimeter-large high-quality Si-doped crystals. We and other authors [24] only detect it in spontaneously-grown micron-sized crystallites that typically contain abundant structural defects such as vacancies. This re-assignment also agrees with theoretical studies predicting that $Si_NV_B$ forms a stable deep center with a ZPL at 0.9 eV [26,30].

### 3.6. Interstitial nature of the BN1 center

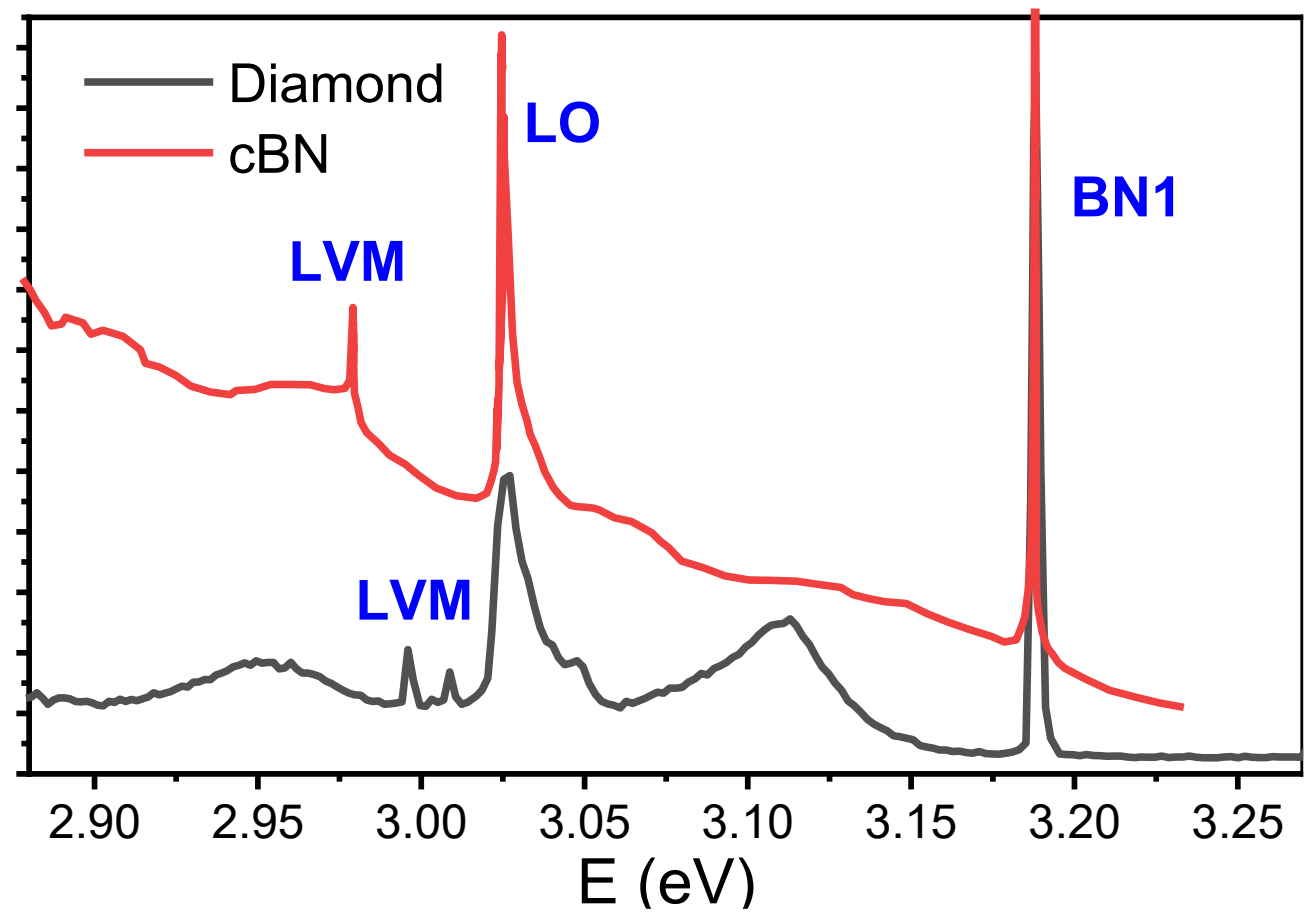

*Figure 6. UV-excited luminescence (77 K) spectra of electron-irradiated diamond and cBN [31]. The cBN spectrum is red-shifted by 106 meV to match the ZPLs and compare the phonon sidebands. LO stands for longitudinal optical and LVM for local vibrational modes.*

HPHT cBN irradiated by high-energy electrons (300 kV, ~$10^{20}$ $e^-/cm^2$) exhibit the BN1 center with a ZPL at 3.294 eV and unusual fine structure [31], with a much wider peak separation as compared to the spin-orbit splitting related to a transition metal (ca. 160 meV vs. 3 meV). In Fig. 6 we compare that structure with the spectrum of the 3.188 eV center in electron-irradiated HPHT diamond (1.5 MeV, ~$10^{18}$ $e^-/cm^2$). For presentation purposes, we downshift the cBN ZPL by 106 meV. A remarkable similarity is observed between the two spectra: both couple to a sharp longitudinal optical (LO) phonon mode with an asymmetric shape. This mode shows an abrupt cut-off at 162-165 meV from the ZPL that correspond to the maximum phonon energy in diamond/cBN (see

Table 1). Both centers also exhibit coupling to a sharp phonon mode with energy beyond the lattice phonon spectrum [32]. Such modes were identified with local molecular-like vibrations of interstitial atoms. Modeling of the mode frequency and its isotopic shift allowed to identify the 3.188 eV luminescence with a complex of substitutional nitrogen atom and a carbon interstitial in diamond [33]. By analogy, we tentatively assign the BN1 center in cBN with an intrinsic interstitial-related defect. This assignment agrees with theory, which predicts a ZPL of ~3 eV for interstitials in cBN [34].

## 4. Conclusions

We systematically compared optical absorption and luminescence in HPHT-grown cBN and diamond crystals synthesized in the same cubic press. This unique experimental approach reveals that many optical centers in cBN have clear analogues in diamond, with their spectral positions blue-shifted by 40–140 meV in accordance with the larger band gap of cBN. By leveraging the well-established defect assignments in diamond, we propose a coherent set of tentative identifications for several key cBN centers.

Our results indicate that the yellow coloration of cBN arises from substitutional oxygen $O_N$, supported by elemental mapping showing oxygen enrichment in yellow regions. The RC1 and RC3 centers exhibit spectral shapes, phonon coupling, and photochromic behavior closely paralleling the $NV^0$ and $NV^-$ centers in diamond, suggesting that they originate from $O_N$–$V_B$ complexes in the neutral and negative charge states. The 1.816 eV luminescence line is reassigned to a $Si_N$–$V_B$ split-vacancy defect, consistent with the behavior of the SiV center in diamond and with the observation that Si-related emission appears only in defect-rich cBN microcrystallites. Finally, the BN1 center shows phonon coupling characteristic of interstitial-related complexes, analogous to the 3.188 eV interstitial-nitrogen defect in diamond.

Together, these findings demonstrate that a cross-material analogy framework provides powerful insight into the defect physics of cBN, where direct identification remains challenging. Future progress will depend critically on the ability to synthesize cBN with controlled impurity concentrations—particularly oxygen—and to systematically introduce and anneal vacancies and interstitials. Such control will enable the full mapping of cBN's defect landscape and may open pathways toward engineered color centers in this material, such as RC1-3 and GC1, for photonic and quantum applications.

## Acknowledgments

CzechNanoLab project LM2023051 funded by MEYS CR is gratefully acknowledged for the financial support of the measurements at CEITEC Nano Research Infrastructure. K.I. is grateful to Prof. G. Davies for electron irradiation.